\documentclass[a4paper,10pt]{article}
\usepackage[fleqn]{amsmath}
\usepackage{amsfonts}
\usepackage{amssymb}
\usepackage{color}

\title{Lorentz Symmetry Breaking in $\mathcal{N} =2$ Superspace }
\author{Mir Faizal$^1$ and Prince A  Ganai$^2$ 
\\ $^1$Department of Physics and Astronomy,  University of Waterloo, \\   Waterloo,
Ontario N2L 3G1, Canada \\
$^2$ Department of Physics, National Institute of Technology, \\
Srinagar, Kashmir-190006, India
}
\date{}
\begin{document}

\maketitle
\begin{abstract} 
  In this paper,   we will study the deformation of a 
three dimensional theory with  $\mathcal{N} =2$  supersymmetry. 
This theory will be deformed by the presence of a constant vector field. This deformation will break the Lorentz 
symmetry. So, we 
  will analyse this theory using    $\mathcal{N} =2$ aether superspace.
The $\mathcal{N} =2$ aether superspace will be obtained from a deformation of the usual $\mathcal{N} =2$   superspace.
This will be done by deforming the generators of the three dimensional  $\mathcal{N} =2$   supersymmetry.
After analysing this deformed superalgebra,   we will derive 
an explicit expression for the superspace propagators in this deformed superspace. 
Finally, we will use these propagators for performing perturbative calculations.  
\end{abstract}
\section{Introduction}
Lorentz symmetry is one of the most important symmetries in nature. However, there are strong theoretical 
indications from various approaches to quantum gravity, that this might only be an effective symmetry 
in nature. Initially the study in this area was motivated by developments in string theory. 
This is because when perturbative string vacuum is unstable, Lorentz symmetry will be naturally broke 
\cite{1}-\cite{2}. 
This is because in this case certain tensors acquire non-zero vacuum expectation value and this introduces a
preferential direction in spacetime. There is a deep relation between string theory and noncommutativity and 
this can also lead to breaking of Lorentz symmetry \cite{01}-\cite{02}.
In fact, the breaking of Lorentz symmetry at Planck scale is generally expected to arise in most theories of 
quantum gravity \cite{04}. 
Furthermore, even though gravity is not renormalizable, it can be made 
renormalizable by adding higher order curvature invariants to the original gravitational Lagrangian \cite{4}. 
However, this spoils the unitarity of the resultant theory \cite{5}. One way out of this problem is to take a 
different Lifshitz scaling for space and time and thus add terms containing higher order spatial derivatives 
without adding any term containing higher order temporal derivative. 
This approach to quantum gravity is called Horava-Lifshitz gravity and in it Lorentz invariance is broken 
in the high energy limit of the theory 
\cite{6}-\cite{7}. Lorentz symmetry breaking has also been studied in the context of
loop quantum gravity \cite{8}-\cite{9}. 
Lorentz symmetry breaking has also been studied in the context of modified dispersion relations 
 and this approach has led to the development of  doubly special special relativity theories
\cite{10}-\cite{a1}. In these theories both the  velocity of light and the Planck energy are
invariant quantities. This assumption naturally 
 incorporates the existence of a
maximal momentum and modifies the first quantized field theory. 
This modification for General Relativity has also been studied and this has led to the development of 
 Gravity's Rainbow \cite{n1}-\cite{n2}. 
Furthermore, the Lorentz symmetry breaking can be used as a possible way to solve the problem of time 
in quantum gravity \cite{gq}. 
 Hence, there are various motivations to study Lorentz symmetry breaking. 
 
  The Lorentz symmetry breaking has also been  
studied in supersymmetric theories \cite{ss}-\cite{ss12}. 
In fact, it is possible to analyse  Lorentz symmetry violation in which a sub-group of the Lorentz 
group is preserved. Thus, for example  Lorentz symmetry can be violated without violating a three-dimensional
rotation subgroup by choosing a background timelike vector field. Such theories have been studied in detail and are 
called aether theories \cite{as}-\cite{as1}. These theories have also been applied in the study of 
Lorentz symmetry violating models of electrodynamics and in these models a
Carroll-Field Jackiw term is added to the original Lagrangian \cite{q1}. 
This term arises as a quantum correction if a Lorentz violating axial term is included in the fermionic sector of the 
original Lagrangian \cite{q2}. This term also breaks the CPT symmetry \cite{q}. It is natural to try to study the 
  supersymmetric theories in aether superspace. In fact, Lorentz symmetry breaking can be inc operated by 
deforming the structure of the generators of supersymmetry and this in turn modifies there superalgebra \cite{l1}. 
Then superderivatives can be constructed such that they anticommute with these modified generators of supersymmetry. 
Some attempts to implement this approach at tree level has also been made \cite{l2}. In fact,  Lagrangian for 
supersymmetric 
scalar field theory with $\mathcal{N} =1$ supersymmetry   has been constructed using eather superspace \cite{uuas}. 
Furthermore,   Lagrangian for       $\mathcal{N} =1$ abelian gauge theories has also been  
constructed using eather superspace \cite{uuas1}. 
In this paper, we will extend this work and study a
supersymmetric field theory with $\mathcal{N} =2$ in aether superspace. 
 We will also
obtain explicit expression for propagators for this theory
and use them for performing perturbative calculations.

\section{$\mathcal{N} = 2$ Aether Superspace}
In this section we will study aether superspace formalism for three dimensional theories. 
We will perform the calculations using $\mathcal{N} = 2$   superspace formalism in three dimensions. 
In order to do that we will 
first consider a constant vector field 
$v^\mu = (v^0, v^i)$, such that,  $||v||^2 = v^\mu v_\mu$. Now
 $||v||^2 =1$ for space-like, $||v||^2 =-1$ for time-like and $||v||^2 =0$ for light-like 
cases \cite{uuas}.   Furthermore,  this constant vector field can be used to construct a
tensor field $k_{\mu\nu} = \alpha v_\mu v_\nu$, for a arbitrary parameter $\alpha$. 
It may be noted that in the  space-like case,   we have   $E^2=p^i p_i+m^2+(2\alpha+\alpha^2)v^i  p_i v^j p_j $.  
So, the  dynamics can be consistently define for, 
$\alpha>0$ and for $\alpha<0$, if  $|\alpha|<< 1$. However, for 
 $\alpha <0$ the theory turns out to be degenerate or unstable. 
In the time-like case, we have $E^2 (1-\alpha^2) = p^i p_i +m^2 $, and so the dynamics is consistent
for all values of $\alpha $,  except  $\alpha =1$. Finally, for the light-like case, we have 
$E(1- 2 \alpha) = [ -2 \alpha \sqrt{p^i p_i} \pm \sqrt{p^i p_i (1+ 2 \alpha + 4 \alpha^2) + m^2} ]$. 
So, the dynamics is consistent for $\alpha << 1$. 

Now  the supersymmetry can be deformed using this vector field, 
in such a way that Lorentz symmetry is broken without breaking any supersymmetry. 
Thus, we can construct two supercharges in three dimensions,
\begin{eqnarray}
 Q_{1a} &=& \partial_{1a} -(\gamma^\mu \partial_\mu\theta_1)_{a} 
- (\gamma^\mu k_{\mu\nu}\partial^\nu\theta_1)_{a}, \nonumber \\
Q_{2a} &=& \partial_{2a} -(\gamma^\mu \partial_\mu\theta_2)_{a} 
- (\gamma^\mu k_{\mu\nu}\partial^\nu\theta_2)_{a}.
\end{eqnarray}
 Now these supercharges satisfy the following superalgebra, 
\begin{eqnarray}
 \{ Q_{1a}, Q_{1b}\} = 2(\gamma^\mu \partial_\mu)_{ab} 
+ 2(\gamma^\mu k_{\mu\nu}\partial^\nu)_{ab}, && \{ Q_{1a}, Q_{2b}\} = 0,\nonumber \\ 
\{ Q_{2a}, Q_{2b}\} = 2(\gamma^\mu \partial_\mu)_{ab} 
+ 2(\gamma^\mu k_{\mu\nu}\partial^\nu)_{ab}. &&
\end{eqnarray}
We can construct  superderivatives which commute with these generators of $\mathcal{N} =2$ supersymmetry,
$\{D_{1a}, Q_{1a}\}= \{D_{2a}, Q_{1a}\} =\{D_{1a}, Q_{2a}\}= \{D_{2a}, Q_{2a}\} =0$. These superderivatives
can be written as 
\begin{eqnarray}
D_{1a} &=& \partial_{1a} + (\gamma^\mu \partial_\mu\theta_1)_{a} 
+ (\gamma^\mu k_{\mu\nu}\partial^\nu\theta_1)_{a}, \nonumber \\
D_{2a} &=& \partial_{2a} + (\gamma^\mu \partial_\mu\theta_2)_{a} 
+ (\gamma^\mu k_{\mu\nu}\partial^\nu\theta_2)_{a}.
\end{eqnarray}
and they satisfy, 
\begin{eqnarray}
 \{D_{1a}, D_{1b}\} = -2(\gamma^\mu \partial_\mu)_{ab} 
-2 (\gamma^\mu k_{\mu\nu}\partial^\nu)_{ab}, && \{ D_{1a}, D_{2b}\}=0,  \nonumber \\  
\{ D_{2a}, D_{2b}\}= -2(\gamma^\mu \partial_\mu)_{ab} 
-2 (\gamma^\mu k_{\mu\nu}\partial^\nu)_{ab}. &&
\end{eqnarray}

We can represent any supersymmetric theory containing two superderivatives $D_{1a}$ and $D_{2a}$ equivalent by two other derivatives  
which are linear combinations of these original superderivatives,
 \begin{eqnarray}
\begin{pmatrix}D_{3a} \\ D_{4a} \end{pmatrix} &=& 
 \begin{pmatrix}x_{11} & x_{12}\\
x_{21} & x_{22}  \end{pmatrix}\begin{pmatrix} D_{1a}\\ D_{2a}\end{pmatrix}, 
 \end{eqnarray}
 where $x_{ij}$ are c-numbers such that, $x_{11}x_{22} - x_{12} x_{21} \neq 0$, so that 
 $D_{3a}$ and $ D_{4a}$ form a valid representation of the supersymmetry.   
For a supersymmetric field theory, the  Jacobian of this transformation can be absorbed in field redefinition. 
Furthermore, it may be noted that as $D_{3a}$ and $D_{4a}$ are linear combinations of $D_{1a}$ and $D_{2a}$, so 
they will also contain $k_{\mu\nu}$ dependent terms.
It may be noted that it also is possible to  analyse a  non-trivially mixing of these supersymmetric 
derivatives  \cite{nactd}.    
 Now we will use a specific form of this transformation, such that \cite{ab12}-\cite{12ab}
 \begin{eqnarray}
  \theta_{ a} = \frac{1}{\sqrt{2}} [\theta_{1a} + i \theta_{2a}  ], &&
 \bar  \theta_{ a} =  \frac{1}{\sqrt{2}} [\theta_{1a} - i \theta_{2a}  ]. 
 \end{eqnarray}  
So, the  derivative $D_a$ and $   \bar D_a $   as 
 \begin{eqnarray}
  D_{a} = \partial_{a} + i(\gamma^\mu \partial_\mu\bar \theta)_{a} 
+ i(\gamma^\mu k_{\mu\nu}\partial^\nu\bar\theta)_{a}, \nonumber \\ 
\bar D_{a} = \bar \partial_{a} + i(\gamma^\mu \partial_\mu\theta)_{a} 
+ i(\gamma^\mu k_{\mu\nu}\partial^\nu\theta)_{a}.
 \end{eqnarray}  
These superderivatives satisfy
\begin{eqnarray}
 \{D_a, \bar D_b \} =   2i(\gamma^\mu \partial_\mu )_{ab} 
+ 2 i(\gamma^\mu k_{\mu\nu}\partial^\nu )_{ab}, && \{\bar D_a, \bar D_b\}= 0, \nonumber \\ 
\{D_a, D_b\} = 0. &&
\end{eqnarray}
We can now construct two supercharges $Q_a$ and $\bar Q_a$, such that 
these superderivatives commute with them, $\{Q_a, D_a\} = \{Q_a, \bar D_a\} =
\{\bar Q_a, D_a\} = \{\bar Q_a, \bar D_a\}  =0$. These supercharges also can be use to represent 
$\mathcal{N} =2$ supersymmetry in three dimensions. 
We can represent them as follows, 
\begin{eqnarray}
 Q_{a} = -i \partial_{a} -(\gamma^\mu \partial_\mu\bar \theta)_{a} 
-(\gamma^\mu k_{\mu\nu}\partial^\nu\bar\theta)_{a}, \nonumber \\ 
\bar Q_{a} = i\bar \partial_{a} + (\gamma^\mu \partial_\mu\theta)_{a} 
+ (\gamma^\mu k_{\mu\nu}\partial^\nu\theta)_{a}.
\end{eqnarray}
They satisfy 
 \begin{eqnarray}
 \{Q_a, \bar Q_b \} =  - 2i(\gamma^\mu \partial_\mu)_{ab} 
- 2 i(\gamma^\mu k_{\mu\nu}\partial^\nu )_{ab}, && \{Q_a, Q_b\} = 0,\nonumber \\
\{\bar Q_a, \bar Q_b\} = 0.  &&
\end{eqnarray}
\section{Superfield Theory in Aether Superspace}
In the previous section we analysed the   supersymmetric algebra for three dimensional aether superspace with $\mathcal{N} =2$
supersymmetry. In this section we will analyse the supersymmetric field theory 
for a three dimensional theory with $\mathcal {N} =2$ supersymmetric in aether superspace. 
We can now define  projections of a $\mathcal{N} =2$ superfield in this aether superspace as 
$D_a \bar \Phi (y, \bar\theta) =0$ and $\bar D_a \Phi (y, \theta) =0$, where $y^\mu = x^\mu + i \theta \bar \theta  \gamma^\mu$.  
Now we expand  these superfields as 
\begin{eqnarray}
 \Phi &=&\phi(y)+\sqrt{2}\theta \psi(y)+\theta^2f(y)
 \nonumber \\ &=& 
 \phi(x) + \sqrt{2} \theta \psi(x) + i \theta \bar \theta \gamma^\mu \partial_\mu \phi(x)
+ i \gamma^\mu \theta \bar \theta k_{\mu\nu}\partial^\mu \phi (x)  \nonumber \\ &&
+ \frac{i}{\sqrt{2}} \theta^2 \bar \theta \gamma^\mu 
\partial_\mu \psi(x) + \frac{i}{\sqrt{2}} \theta^2 \bar \theta \gamma^\mu k_{\mu\nu}
\partial^\nu \psi (x) -\frac{1}{4}  \theta^2 \bar \theta^2 \partial^\mu \partial_\mu \phi (x)
\nonumber \\ &&- \frac{1}{2} \theta^2 \bar \theta^2 
\partial^\mu \partial^\nu k_{\mu\nu}\phi(x) -\frac{1}{4}  \theta^2 \bar \theta^2 
 k^{\tau\mu } k_{\tau\nu} \partial_\mu \partial^\nu \phi(x) + \theta^2 f(x), 
 \nonumber \\ 
 \bar\Phi &=& \phi^*(y)+ \sqrt{2}\bar \theta\bar\psi(y)+\bar \theta^2f^*(y)
 \nonumber \\&=&
 \phi^* (x)+ \sqrt{2}\bar \theta \bar\psi(x) + i \theta \bar \theta \gamma^\mu \partial_\mu \phi^*(x)
+ i \gamma^\mu \theta \bar \theta k_{\mu\nu}\partial^\mu \phi^* (x) 
\nonumber \\ && + \frac{i}{\sqrt{2}} \bar \theta^2  \theta \gamma^\mu 
\partial_\mu \bar \psi (x)+ \frac{i}{\sqrt{2}}\bar  \theta^2 \theta \gamma^\mu k_{\mu\nu}
\partial^\nu \bar \psi(x)  -\frac{1}{4} \theta^2 \bar \theta^2 \partial^\mu \partial_\mu \phi^* (x)
\nonumber \\ && -  \frac{1}{2} \theta^2 \bar \theta^2 
\partial^\mu \partial^\nu k_{\mu\nu}\phi^* (x) - \frac{1}{4} \theta^2 \bar \theta^2 
 k^{\tau\mu } k_{\tau\nu} \partial_\mu \partial^\nu \phi^*(x) + \bar \theta^2 f^* (x). 
\end{eqnarray}
Now we can write the action for a supersymmetric field theory in this aether superspace as follows,  
\begin{equation}
 S = \frac{1}{2} \int d^3x [2d^2 \theta d\bar \theta^2 \bar \Phi \Phi + m d^2 \theta \Phi^2  + m d^2 \bar  \theta \bar \Phi^2 ]. 
\end{equation}
This action can be written in component form as, 
\begin{eqnarray}
 S=\int d^3 x\ \bigg[ \begin{pmatrix}\phi^* & f  \end{pmatrix}
 \begin{pmatrix}-  \partial^\mu \partial_\mu  - 2 \partial^\mu \partial^\nu k_{\mu\nu} -
 k^{\tau\mu } k_{\tau\nu} \partial_\mu \partial^\nu & m\\
 m & 1  \end{pmatrix}\begin{pmatrix} \phi \\ f^* \end{pmatrix} \nonumber \\
 +\frac{1}{2}\begin{pmatrix}\bar{\psi} & {\psi}\end{pmatrix}
\begin{pmatrix} -i\gamma^\mu\partial_\mu -i \gamma^\mu k_{\mu \nu}\partial^\nu & -m \\-m & -i\gamma^\mu\partial_\mu 
-i \gamma^\mu k_{\mu \nu}\partial^\nu \end{pmatrix}\begin{pmatrix} \psi \\ \bar\psi \end{pmatrix}\bigg]. 
\end{eqnarray}
Now in general for any field, the generating functional  for the Green's functions is given by 
\begin{equation}
 Z[J,J^*]=N\exp {-i\int d^3 x JK^{-1}J^*},
\end{equation}
where $N$ is normalization constant. Thus, the Green's function for any field can be written as $iK^{-1}$. 
The $K_B$ for the bosonic part is given by 
\begin{equation}
  K_B=\begin{pmatrix} -\partial^\mu \partial_\mu  - 2 \partial^\mu \partial^\nu k_{\mu\nu} -
 k^{\tau\mu } k_{\tau\nu} \partial_\mu \partial^\nu & m \\ m & 1 \end{pmatrix},  
 \end{equation}
 which can be inverted to obtain, 
 \begin{eqnarray}
K_B^{-1}&=&\frac{1}{-\partial^\mu \partial_\mu  - 2 \partial^\mu \partial^\nu k_{\mu\nu} -
 k^{\tau\mu } k_{\tau\nu} \partial_\mu \partial^\nu -m^2}\nonumber\\
\nonumber\\ &&
\times \begin{pmatrix}1 & -m \\ -m & -\partial^\mu \partial_\mu  - 2 \partial^\mu \partial^\nu k_{\mu\nu} -
 k^{\tau\mu } k_{\tau\nu} \partial_\mu \partial^\nu \end{pmatrix}. 
\end{eqnarray}
Similarly, $K_F$ for the fermionic part is given by  
\begin{equation}
 K_F=\begin{pmatrix} -i\gamma^\mu\partial_\mu -i \gamma^\mu k_{\mu \nu}\partial^\nu
 & -m \\ -m &-i\gamma^\mu\partial_\mu  -i \gamma^\mu k_{\mu \nu}\partial^\nu
\end{pmatrix},
\end{equation}
which can be inverted to obtain, 
\begin{eqnarray}
K_F^{-1}&=& \frac{1}{-\partial^\mu \partial_\mu  - 2 \partial^\mu \partial^\nu k_{\mu\nu} -
 k^{\tau\mu } k_{\tau\nu} \partial_\mu \partial^\nu -m^2}\nonumber\\ \nonumber\\ && \times
 \begin{pmatrix} -i \gamma^\mu\partial_\mu - \gamma^\mu k_{\mu \nu}\partial^\nu & m\\
                                            m & -i\gamma^\mu\partial_\mu -i\gamma k_{\mu \nu}\partial^\nu
                                            \end{pmatrix}.
\end{eqnarray}
Now using $ K_B^{-1}$ and $K_{F}^{-1}$, we can calculate the two-point functions for all the component fields in this theory, 
\begin{eqnarray}
 <0|\phi(x)\phi^*(x^\prime)|0> & =&\frac{i}{-\partial^\mu \partial_\mu  - 2 \partial^\mu \partial^\nu k_{\mu\nu} -
 k^{\tau\mu } k_{\tau\nu} \partial_\mu \partial^\nu-m^2}\nonumber \\ && \times \delta^3  (x-x^\prime), \nonumber \\
 <0|\phi(x)f(x^\prime)|0> & =&\frac{-im}{-\partial^\mu \partial_\mu  - 2 \partial^\mu \partial^\nu k_{\mu\nu} -
 k^{\tau\mu } k_{\tau\nu} \partial_\mu \partial^\nu)-m^2}\nonumber \\ && \times \delta^3  (x-x^\prime), \nonumber\\
 <0|\phi^*(x)f^*(x^\prime)|0> & =&\frac{-im}{-\partial^\mu \partial_\mu  - 2 \partial^\mu \partial^\nu k_{\mu\nu} -
 k^{\tau\mu } k_{\tau\nu} \partial_\mu \partial^\nu-m^2}\nonumber \\ && \times \delta^3  (x-x^\prime),\nonumber \\
 <0|f(x)f^*(x^\prime)|0> & =&\frac{-i(\partial^\mu \partial_\mu  + 2 \partial^\mu \partial^\nu k_{\mu\nu} +
 k^{\tau\mu } k_{\tau\nu} \partial_\mu \partial^\nu ) }{-\partial^\mu \partial_\mu  - 2 \partial^\mu \partial^\nu k_{\mu\nu}-
 k^{\tau\mu } k_{\tau\nu} \partial_\mu \partial^\nu-m^2}\nonumber \\ && \times \delta^3  (x-x^\prime),\nonumber\\
 <0|\psi_a (x)\bar{\psi}_{b} (x^\prime)|0> & =&\frac{\gamma^{\mu}_{ab}(\partial_\mu + k_{\mu \nu}\partial^\nu)}
 {-\partial^\mu \partial_\mu  - 2 \partial^\mu \partial^\nu k_{\mu\nu} -
 k^{\tau\mu } k_{\tau\nu} \partial_\mu \partial^\nu -m^2}\nonumber \\ && \times \delta^3  (x-x^\prime), \nonumber \\
  <0|\psi_a(x)\psi_b(x^\prime)|0> & =&\frac{im\delta_{ab}}{-\partial^\mu \partial_\mu  - 2 \partial^\mu \partial^\nu k_{\mu\nu} -
 k^{\tau\mu } k_{\tau\nu} \partial_\mu \partial^\nu-m^2} \nonumber \\ && \times \delta^3  (x-x^\prime),\nonumber  \\
 <0|\bar{\psi}_{a}(x) \bar{\psi}_{b}(x^\prime)|0> & =&\frac{im\delta_{ab}}{-\partial^\mu \partial_\mu  -
 2 \partial^\mu \partial^\nu k_{\mu\nu} -
 k^{\tau\mu } k_{\tau\nu} \partial_\mu \partial^\nu-m^2}\nonumber \\ && \times \delta^3  (x-x^\prime).
\end{eqnarray}
Using these two-point functions, and the fact that 
 $  <0|\Phi(y,\theta)\Phi(y^\prime,\theta^\prime)|0> 
 = \theta^{\prime 2}<0|\phi(y)f(y^\prime)|0> +\theta^2<0|f(y)\phi(y^\prime)|0>+2\theta^a\theta^{\prime b} <0|\psi_a(y)\psi_b (y^\prime)|0> $, 
 we obtain 
\begin{equation}
  <0|\Phi(y,\theta)\Phi(y^\prime,\theta^\prime)|0>=
\frac{[-im(\theta^{\prime 2}+\theta^2) +2\theta\theta^\prime im]
\delta^3 (y-y^\prime)}{-\partial^\mu \partial_\mu  - 2 \partial^\mu \partial^\nu k_{\mu\nu} -
 k^{\tau\mu } k_{\tau\nu} \partial_\mu \partial^\nu-m^2}. 
 \end{equation}
 Now we  write $-im(\theta^{\prime 2}+\theta^2) +2\theta\theta^\prime im 
 = -im(\theta-\theta^\prime)^2 \delta^3 (y-y^\prime) $,  
and use the identity, 
$4 (\theta-\theta^\prime)^2F(y-y^\prime) = - (\theta-\theta^\prime)^2[\bar{D}^2(\bar{\theta}-\bar{\theta^\prime})^2]F(y-y^\prime) 
 =- \bar{D}^2[(\theta-\theta^\prime)^2(\bar{\theta}-\bar{\theta^\prime})^2F (y-y^\prime)] 
 =- \bar{D}^2[(\theta-\theta^\prime)^2(\bar{\theta}-\bar{\theta^\prime})F(x-x^\prime)]$, 
to obtain the following expression,
\begin{equation}
 <0|\Phi(x,\theta,\bar{\theta})\Phi(x^\prime,\theta^\prime,\bar{\theta^\prime})|0>
 =\frac{im}{4}\bar{D}^2 \mathcal{M}_1 (x, \theta, \bar \theta), 
 \end{equation}
 where 
 \begin{equation}
  \mathcal{M}_1 (x, \theta, \bar \theta) = \frac{\delta^2(\theta-\theta^\prime)\delta^2(\bar{\theta}-
 \bar{\theta^\prime})\delta^3 (x-x^\prime) }{-\partial^\mu \partial_\mu  - 2 \partial^\mu \partial^\nu k_{\mu\nu}-
 k^{\tau\mu } k_{\tau\nu} \partial_\mu \partial^\nu-m^2}.
 \end{equation}
Similarly, using  the fact that   $  <0|\bar \Phi(y,\bar \theta)\bar \Phi(y^\prime,\bar \theta^\prime)|0> 
 = \bar \theta^{\prime 2}<0|\phi^*(y)f^*(y^\prime)|0> +\bar \theta^2<0|f^*(y)\phi^*(y^\prime)|0>
 +2\bar \theta^a\bar \theta^{\prime b} <0|\bar \psi_a(y)\bar \psi_b (y^\prime)|0> $, we obtain 
 \begin{equation}
 <0|\bar{\Phi}(x,\theta,\bar{\theta})\bar{\Phi}(x^\prime,\theta^\prime,\bar{\theta^\prime})|0>= \frac{im}{4}D^2
 \mathcal{M}_2 (x, \theta, \bar \theta),
 \end{equation}
 where
\begin{equation}
  \mathcal{M}_2 (x, \theta, \bar \theta) =\frac{\delta^2(\theta-\theta^\prime)\delta^2(\bar{\theta}-\bar{\theta^\prime})\delta^3
 (x-x^\prime)}{-\partial^\mu \partial_\mu  - 2 \partial^\mu \partial^\nu k_{\mu\nu} -
 k^{\tau\mu } k_{\tau\nu} \partial_\mu \partial^\nu -m^2}.
\end{equation}
Finally, using  the fact that   $  <0| \Phi(y, \theta)\bar \Phi(y^\prime,\bar \theta^\prime)|0> 
 = \theta' \bar \theta^{\prime }<0|\phi (y)f^*(y^\prime)|0> +\theta \bar \theta <0|f (y)\phi^*(y^\prime)|0>
 +2\theta^a\bar \bar \theta^{\prime b} <0|  \psi_a(y)\bar \psi_b (y^\prime)|0> $, we obtain

\begin{equation}
 <0|\bar{\Phi}(x,\theta,\bar{\theta})\bar{\Phi}(x^\prime,\theta^\prime,\bar{\theta^\prime})|0> =
 \frac{  \delta^3 (y-y')- \mathcal{N} (\theta, \theta' )\delta^3 (y-y')}{-\partial^\mu \partial_\mu 
 - 2 \partial^\mu \partial^\nu k_{\mu\nu}-
 k^{\tau\mu } k_{\tau\nu} \partial_\mu \partial^\nu-m^2}, 
\end{equation}
where we have defined,  $  \mathcal{N} ( \theta, \theta' )
=2i\theta^a\bar{\theta^\prime}^{b } i\gamma^\mu_{ab}
 (\partial_\mu + k_{\mu \nu}\partial^\nu) 
+ i\theta^2\bar{\theta^{2\prime} } ( \partial^\mu \partial_\mu 
 + 2 \partial^\mu \partial^\nu k_{\mu\nu}
 + k^{\tau\mu } k_{\tau\nu} \partial_\mu \partial^\nu)$. 
This can be simplified to   obtain the following expression  
\begin{equation}
 <0|\Phi(x,\theta,\bar{\theta})\bar{\Phi}(x^\prime,\theta^\prime,\bar{\theta^\prime})|0> = 
 \frac{i}{16}\bar{D}^2 {D^\prime}^2 \mathcal{M}_3 (x, \theta, \bar \theta)  , \label{prop3}
\end{equation}
where 
\begin{equation}
 \mathcal{M}_3(x, \theta, \bar \theta)  =  \frac{\delta^2(\theta-\theta^\prime)\delta^2(\bar{\theta}-\bar{\theta^\prime})
 \delta^3 (x-x^\prime)}{- \partial^\mu \partial_\mu  - 2 \partial^\mu \partial^\nu k_{\mu\nu} -
 k^{\tau\mu } k_{\tau\nu} \partial_\mu \partial^\nu -m^2}. 
\end{equation}
\section{Interactions in Aether Superspace}
These superspace propagator's can now be used for analyzing superspace perturbations.   
So,  we can calculate  the  loop correction to the superspace propagator's for different 
interaction terms.    
We can start by calculating the one-loop corrections for 
$<0|\Phi(x,\theta,\bar{\theta}) {\Phi}(x^\prime,\theta^\prime, {\theta^\prime})|0>$, when the interaction term  
is  of the form $\mathcal{L}_{int} = \lambda  d^2 \theta  \Phi^3/3$. 
The one-loop corrections to 
$<0|\Phi(x,\theta,\bar{\theta}) {\Phi}(x^\prime,\theta^\prime, {\theta^\prime})|0>  $, 
can be written as 
\begin{eqnarray}
&& -2\lambda^2 \left(\frac{im}{4}\right)^4\int d^3 p_2d^2\theta_1d^2\theta_2
 \frac{\bar{D}^2_1\delta^2(\theta_1-\theta_2)\delta^2(\bar{\theta}_1-\bar{\theta}_2)}{{\cal A}(p_1-p_2,k,m)}
 \nonumber \\ && \times 
  \frac{\bar{D}^2\delta^2(\theta-\theta_1)\delta^2(\bar{\theta}-\bar{\theta}_1) }{ {p_1}^2 
  +2k_{\mu \nu}{p_1}^{\mu} {p_1}^{\nu} + k^{\tau\mu} k_{\tau\nu} {p_1}_{\mu} {p_1}^{\nu}-m^2}
 \nonumber\\ &&\times 
  \frac{ \bar{D}^2_1
  \delta^2(\theta_1-\theta_2)\delta^2(\bar{\theta}_1-\bar{\theta}_2)
   }{ {p_2}^2 +2k_{\mu \nu}{p_2}^{\mu} {p_2}^{\nu} 
  + k^{\tau\mu} k_{\tau\nu} {p_2}_{\mu} {p_2}^{\nu}-m^2}
\nonumber\\ &&
  \times  \frac{\bar{D^\prime}^2\delta^2(\theta_2-\theta^\prime)\delta^2(\bar{\theta_2}-\bar{\theta^\prime})}{
  {p_1}^2 +2k_{\mu \nu}{p_1}^{\mu} {p_1}^{\nu} + k^{\tau\mu} k_{\tau\nu} {p_1}_{\mu} {p_1}^{\nu}-m^2}.
\end{eqnarray}
where ${\cal A}(p_1-p_2,k,m) = (p_1-p_2)^2+2k_{\mu\nu}(p_1-p_2)^\mu(p_1-p_2)^\nu 
  +k^{\tau\mu}k_{\tau\nu}(p_1-p_2)_\mu(p_1-p_2)^\nu-m^2$. This expression vanishes
  due to the odd parity of the superspace coordinates, and so 
  the mass parameter appearing in the superpotential does not get renormalized. 
It may be noted that the  one-loop contributions 
$ <0|\bar{\Phi}(x,\theta,\bar{\theta})\bar{\Phi}(x^\prime,\theta^\prime,\bar{\theta^\prime})  |0>$
will also vanish.  However, the      one-loop contributions   $ 
  <0| {\Phi}(x,\theta,\bar{\theta})\bar{\Phi}(x^\prime,\theta^\prime,\bar{\theta^\prime})  |0>$
  do not vanish. 
  In fact, we can write the one-loop corrections  to this propagator as 
 \begin{eqnarray}
&& 2\lambda\lambda^* \left(\frac{i}{16}\right)^2 \int d^3 p_2 d^2\theta_1d^2\bar{\theta}_2
 \frac{\bar{D}_1^2D_2^2 \delta^2 (\theta_1-\theta_2 )\delta^2 ( \bar{\theta}_1-
\bar{\theta}_2)}{{\cal A}(p_1-p_2,k,m)}\nonumber \\ && \times 
 \frac{ \bar{D}_1^2D_2^2\delta^2 (\theta_1-\theta_2 )\delta^2 ( \bar{\theta}_1-
\bar{\theta}_2)  }{ 
 {p_2}^2 +2k_{\mu \nu}{p_2}^{\mu} {p_2}^{\nu} + k^{\tau\mu} k_{\tau\nu} {p_2}_{\mu} {p_2}^{\nu}-m^2  }. 
\end{eqnarray}
This integral is divergent and will require renormalization of the superfield. 

We can analyse  the loop corrections to the vacuum energy. The 
one-loop   corrections to the vacuum energy also vanish. This is becuase they involve a
two-point function evaluated at the same point, and 
$\delta(\theta - \theta) = 0$ \cite{su}-\cite{s}. 
At two-loops, there is a non-trivial diagram for the vacuum energy. 
\begin{eqnarray}
 && 4\lambda\lambda^* \big(\frac{i}{16}\big)^3 \int d^3  p_1 d^3 
p_2 d^2\theta_1 d^2 \bar{\theta}_2 \frac{\bar{D}^2_{-p_1-p_2}D^2_{-p_1-p_2}\delta^2 (\theta_1-\theta_2 )\delta^2 ( \bar{\theta}_1-
\bar{\theta}_2)}{{\cal A}(p_1+p_2,k,m)}\nonumber \\ && \times 
\frac{\bar{D}^2_{p_1}D^2_{p_1}\delta^2 (\theta_1-\theta_2 )\delta^2 ( \bar{\theta}_1-
\bar{\theta}_2) }{ {p_1}^2 +2k_{\mu \nu}{p_1}^{\mu} {p_1}^{\nu} + k^{\tau\mu} k_{\tau\nu} {p_1}_{\mu} {p_1}^{\nu}-m^2}\nonumber\\
&&
\times\frac{\bar{D}^2_{p_2}D^2_{p_2}\delta^2 (\theta_1-\theta_2 )\delta^2 ( \bar{\theta}_1-
\bar{\theta}_2)}{ ({p_2}^2 +2k_{\mu \nu}{p_2}^{\mu} {p_2}^{\nu} + k^{\tau\mu} k_{\tau\nu} {p_2}_{\mu} {p_2}^{\nu}-m^2)}
\nonumber \\ &=& 
 4\lambda\lambda^*\big(\frac{i}{16}\big)^3 \int d^3  p_1 d^3  p_2 d^2\theta_1d^2\bar{\theta}_2\frac{1}{ {\cal A}(p_1+p_2,k,m)}
 \nonumber \\ && \times 
 \frac{1}{ {p_1}^2 +2k_{\mu \nu}{p_1}^{\mu} {p_1}^{\mu}
 + k^{\tau\mu} k_{\tau\nu} {p_1}_{\mu} {p_1}^{\nu}-m^2}\nonumber\\ && 
\times\frac{1}{ {p_2}^2 +2k_{\mu \nu}{p_2}^{\mu} {p_2}^{\nu} + k^{\tau\mu} k_{\tau\nu} {p_2}_{\mu} 
{p_2}^{\nu}-m^2}
\nonumber \\ &=& 0.  
\end{eqnarray}
So,  even the contributions from this 
non-trivial diagram vanish. 
Thus,  the vacuum energy for the aether superspace is still zero even 
at two-loops. 
It may be noted that the 
quantum fluctuations do not break the supersymmetry in three dimensional $\mathcal{N} =2$ aether superspace. 
It would be interesting to analyse general non-renormalization theorems for the aether superspace.  
\section{Conclusion}
In this paper, we analysed a three dimensional supersymmetric field theory with $\mathcal{N} =2$ supersymmetry in aether superspace. 
In this superspace the Lorentz symmetry was broken without breaking any supersymmetry. 
We analysed this model in a representation where a   mixing between the original generators of 
$\mathcal{N} =2$ supersymmetry occurred. We then obtained an explicit expression for supercharges and superderivatives 
in this representation of 
$\mathcal{N} =2$ supersymmetry. We used these superderivatives in aether superspace to derive explicit expressions for 
propagators for our model. Finally, we used these propagators for performing some perturbative calculations. 
We thus observed that there is no contribution to the 
vacuum energy  from one-loop and   two-loops graphs. It was   argued that 
the supersymmetry is not broken by quantum fluctuations 
in aether superspace, at least till two loops. 
It will be interesting to perform a similar calculation for models with higher amount of supersymmetry. 
Thus, we could analyse a four dimensional scalar superfield model in $\mathcal{N} =2$ aether superspace. 
It will also be interesting to study the Lorentz symmetry breaking by adding CPT odd Lorentz-breaking terms
to the components of superfields and keeping the superalgebra undeformed. Furthermore, we can also add explicit 
Lorentz breaking terms to the superspace action. The action derived from such an approach will contain higher 
derivative terms.

\end{document}